# Transcriptomic Causal Networks identified patterns of differential gene regulation in human brain from Schizophrenia cases versus controls

Akram Yazdani[1], Raul Mendez-Giraldez[2], Michael R Kosorok[3], Panos Roussos[1,4,5]

[1]Department of Genetics and Genomic Science, Icahn School of Medicine at Mount Sinai, New York, NY, USA
[2]Lineberger Comprehensive Cancer Center, School of Medicine, University of North Carolina at Chapel Hill, NC, USA
[3]Department of Biostatistics, University of North Carolina at Chapel Hill, NC, USA
[4]Department of Psychiatry and Friedman Brain Institute, Icahn School of Medicine at Mount Sinai, New York, NY 10029, USA
[5]Mental Illness Research Education and Clinical Center (MIRECC), James J. Peters VA Medical Center, Bronx, New York, 10468, USA

## Abstract

Common and complex traits are the consequence of the interaction and regulation of multiple genes simultaneously, which work in a coordinated way. However, the vast majority of studies focus on the differential expression of one individual gene at a time. Here, we aim to provide insight into the underlying relationships of the genes expressed in the human brain in cases with schizophrenia (SCZ) and controls. We introduced a novel approach to identify differential gene regulatory patterns and identify a set of essential genes in the brain tissue. Our method integrates genetic, transcriptomic, and Hi-C data and generates a transcriptomic-causal network. Employing this approach for analysis of RNA-seq data from CommonMind Consortium, we identified differential regulatory patterns for SCZ cases and control groups to unveil the mechanisms that control the transcription of the genes in the human brain. Our analysis identified modules with a high number of SCZ-associated genes as well as assessing the relationship of the hubs with their down-stream genes in both, cases and controls. In addition, the results identified essential genes for brain function and suggested new genes putatively related to SCZ.

## Introduction

Differences in gene expression are likely to underpin much of human diversity, including psychiatric disorders such as schizophrenia (SCZ). Schizophrenia is a severe, complex, and heritable psychiatric disorder with a world-wide prevalence of 1%, characterized by abnormalities in thought and cognition. Recent studies on SCZ, which comprise linkage scans and their meta-analyses, candidate gene association analyses, differential expression analyses and genome-wide association studies (GWAS), have investigated genes/markers and chromosomal regions for SCZ (Ng et al., 2009; International Schizophrenia Consortium et al., 2009; Shi et al., 2009; Stefansson et al., 2009; Ripke et al., 2014). These studies show that SCZ disorder involves changes in multiple genes, each conferring small and incremental risk. The risks potentially converge in deregulated biological pathways, cellular functions and local circuit that eventually scale up to brain region pathophysiology (Belmaker and Agam, 2008; Sibille and French, 2013). Therefore, understanding the impact of these effects on pathophysiology of SCZ is essential to reveal their relationship as a system. However, typical methods for differential expression analysis are based on the analysis of a single gene at a time regardless of its relation with other genes.

To divulge underlying relationships of genes related to SCZ, we focus on identifying transcriptomic causal networks in an observational study and introduce a novel approach established in Mendelian Randomization and Bayesian graphical modeling. This approach is an adaptation of the genome directed acyclic graph (G-DAG) algorithm (Yazdani et al., 2016) for transcriptomic networks, which was originally proposed for metabolomics. Our method differs from correlation-based networks (He, Chen and Evans, 2007; Wright et al., 1999; Bullmore and Sporns, 2009; Fromer et al., 2016) in that the latter tend to result in many dependencies that arise from undirected networks and face limited success at finding functional genes (Sullivan, Kendler, & Neale, 2003). However, the causal networks are directed graphs established in the recognition of the hierarchal structure of the biological systems, which provide better understanding of the regulatory patterns, as well as their causal interpretation. In addition, the causal network results into novel intervention targets and disease-associated genes with reproducible results. Studying of transcriptomic causal networks leads to the identification of differential gene regulation patterns, unveils the mechanisms that control the transcription of the SCZ-associated genes, and narrows down the search space for finding trans-regulatory elements to a subset of genes. Considering all together, the transcriptomic causal network facilitates efficient experimental study design for further analysis and provides a mechanistic understanding of disease processes (Pearl, 2000).

In the present study, we used RNA-seq data from the dorsolateral prefrontal cortex of cases with SCZ and controls generated as part of the CommonMind Consortium, to build a transcriptomic causal network. Our study shows that genes whose transcription is highly regulated, i.e. they are highly connected in the network, are less likely to be affected by genetic variants. We identified 9 genes differentially regulated in SCZ cases versus controls: *GABRA2, LRRTM2, PPM1E, SORT1, GNAL, ZNF69, RALGPS2, ZNF672, SNRNP48*. Among these genes, *GABRA2, LRRTM2, PPM1E* have preferential transcription in human brain.

Mapping the SCZ-associated genes on the transcriptomic network, we defined modules of genes characterized by a high impact gene ("the hub") on the transcription of downstream genes and then evaluated these effects with penalization and cross-validation methods. The observed pathways are made of a small subset of genes involving trans-regulatory interactions between genes in different chromosomes, as opposed the genetic variants that exert *cis*-regulatory effect on genes in the same chromosome. As a result, three of the identified hubs (*NRXN3, TENM3* and *MYH10*) were essential genes for brain function since that are good targets for new therapeutic treatments since they control the transcription level of 15 genes, in both cases and controls. In addition, we discovered *RNAF150*, *UNC5D*, *RTF*, *LMO7* and *SEZ6L* as novel SCZ-associated.

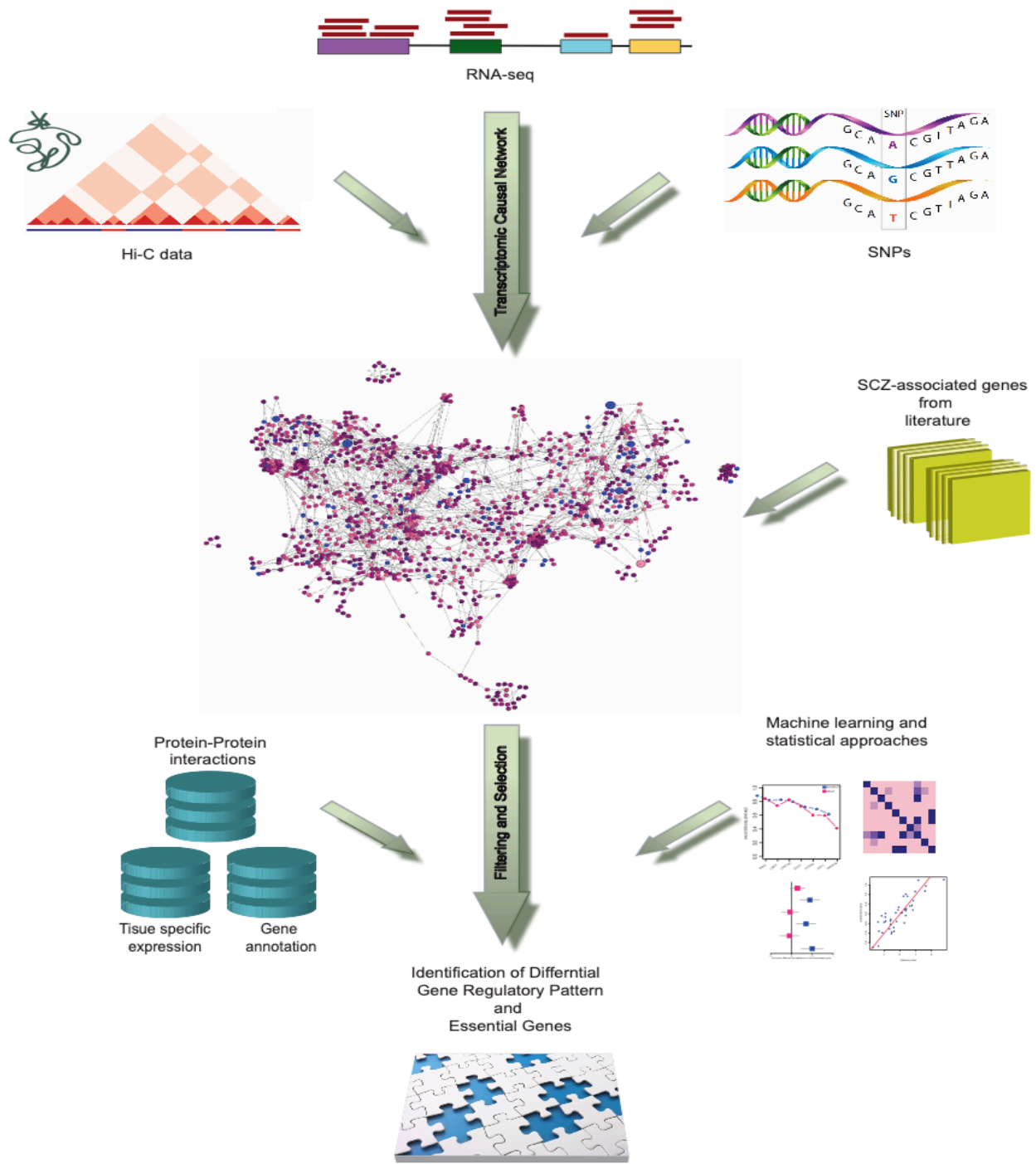

**Figure 1**: Schematic representation of our approach. First it builds a causal network using transcriptomic, genotype and chromatin 3D contact data. Second, it identifies differential regulatory patterns and essential genes.

## Data

### RNA-seq

Study sample: The RNA sequence (RNA-seq) data are from post-mortem dorsolateral prefrontal cortex which controls complex, higher-level cognitive and executive functions. These data are available in CommonMind Consortium (CMC; http://www.synapse.org/CMC). Following data normalization, 16,423 genes (based on Ensemble models) were expressed at levels sufficient for analysis, of which 14,222 were protein-encoding. Validation using PCR showed high correlation (r > 0.5) with normalized transcription from RNA-seq for the majority of genes assessed. Covariates for RNA integrity (RIN), library batch, institution (brain bank), diagnosis, age at death, genetic ancestry, post-mortem interval and sex together explained a substantial fraction (0.42) of the average variance of gene transcription and were thus employed to adjust the data for the analyses (Fromer et al., 2016).

### Genotype data

Samples were genotyped on the Illumina Infinium HumanOmniExpressExome array (958,178 single-nucleotide polymorphisms, or SNPs). These genotypes were used to detect SNPs with influence on gene transcription, to estimate ancestry of the samples and to ensure sample identity across DNA and RNA experiments. The distribution of ethnicities shows Caucasian 80.7%, African American 14.7%, Hispanic 7.7%, and East Asian 0.6%. To prevent the study from being confounded due to different ethnicities, we included 80.7% Caucasian (209 and 206 samples in SCZ control group respectively) in this study.

### Hi-C data

The Hi-C libraries are constructed from mid-gestation developing human cerebral cortex during the peak of neurogenesis and migration from two major zones: the cortical and subcortical plate, consisting primarily of post-mitotic neurons and the germinal zone, containing primarily mitotically active neural progenitors (Won et al., 2016).

## Overview of the method

Figure 1 provides a schematic representation of the approach, while each step is briefly explained in the following and detailed in method section.

### Step1: Building transcriptomic causal network

We first extract genetic information in a new fashion considering that transcription is affected by genetic variations nearby genes (orientation, position, and distance) ( Yazdani, Mendez Giraldez, & Samiei, 2018). For each gene, we select genetic variants that mediate transcription of genes in *cis*-regulatory elements via both short- and long-range interactions using chip exome and high-resolution 3D maps of chromatin (Hi-C). We then adapt G-DAG algorithm (Yazdani et al., 2016) for integrating our genetic variants and then built transcriptomic causal network established in Mendelian principles based on the fact that the genetic inherited variation is the cause of gene transcription not the other way around. We assess the quality of the fit by employing Hamming distance (Tsamardinos, Brown, & Aliferis, 2006) and test the stability of the network using a backward selection techniques.

**Step 2: Identification of receptor, broadcaster, and mediator genes**

The casual network properties measure the genes impacts on the transcription network based on their *out-degree* and *in-degree* of a genes. The former refers to the number of downstream genes directly affected by a gene of interest, whereas the latter refers to the number of upstream genes that affect the corresponding target gene. Based on their connectivity, genes can be classified as broadcaster if they have a given out-degree and zero in-degree; receptors if they have a given in-degree and zero out-degree; and mediator if they have both non-zero out- and in-degrees. For instance, *RAB30* in Figure 2A (closeup), has out-degree of 3, *ZNF770*, *LCOR*, *MAP3K2* and *MOB1B* are the receptor genes and *TYGO1*, *LRRC58*, *SESN3* are mediators. The effect of a gene can be propagated downstream until it reaches a receptor on a given path that is called *pathway*. Genes with high number of out-degree and long pathway across the network are called high impact genes or hubs. We further validated the impact of the genes in the network using penalized method and cross-validation techniques.

**Step 3: Module identification**

We define a module as a subnetwork that includes a high impact gene (i.e. the "hub") and those genes that are highly influenced by the hub directly or indirectly. The boundaries of the modules are made of receptor genes that prevents the effect of the hub to be distributed beyond these genes. In order to identify modules related to SCZ, we mapped the SCZ-associated genes, according to liturature, on the transcriptomic network. Through further analyses, we assess the impact of the hub gene in the modules on the transcription level of non-hub genes using linear regression model and cross-validation technique.

**Step 4: Identifying differential regulatory patterns, essential genes for brain function and novel genes putatively related to SCZ**

To estimate differential patterns between SCZ-cases and controls, we seek genes with high degree of connectivity in the networks whose loss of their connection might have significant effect on brain function. We exclude genes with only one upstream effector and search for differential patterns among the remaining gene subset.

We next assessed the impact of the hub on the transcription of non-hub genes within each module, through linear prediction model and compare SCZ cases versus controls. The hubs that are good predictors for the transcription of downstream genes in both cases and controls are considered as essential for the normal brain function. The genes that are predicted differently between cases and controls correspond to the genes with alternation in their connectivity by acquiring or losing interaction in SCZ versus control groups.

Given the list of SCZ-associated genes that are mapped in our causal network, we found new genes putatively related to SCZ and evaluated our results using conditional analysis assuming that the SCZ-associated genes are truly related to SCZ.

**Step5: Functional annotation of genes and broadcaster – receptor interactions**

To better understand the effect on transcription level exerted by one gene (broadcaster) onto another (receptor), we looked at different databases. Basic functional annotation and the proteins coded for the various genes provided from the UniProt KnowledgeBase (The UniProt Consortium, 2017). The latter was used to check for potential Transcription Factors or Transcription regulators. In many cases, the depicted regulatory relationships corresponded to genes in different chromosomes, i.e. typical trans-interactions, then we further checked the presence of non-coding RNAs overlapping any of the exons of the gene of interest according to the NCBI Genome Browser using NCBI (GRCh38.p13/p12) or Ensemble Genes definition (release 96), available through the Entrez Gene (Maglott, Ostell, Pruitt, & Tatusova, 2005). Since the reads from the RNA-seq experiment were aligned to a much older reference genome, hg19 (Fromer et al., 2016) the counts matching both exons of a given gene and non-coding RNA according to the newer annotations, could well correspond to missannoated non-coding RNA, with potential trans-regulatory effect over distal genes. For each relevant gene within each of the 5 described modules, we accounted for the level of expression in human brain and other relevant tissues according to the Human Protein Atlas (Uhlén et al., 2015), provided also by the Entrez Genes resource (Maglott et al., 2005). Given that it had been earlier determined that co-regulated proteins are likely to interact (Segal, Wang, & Koller, 2003), additional information about possible protein – protein interactions at different degree of evidences, from experiment to co-citations in PubMed abstracts was obtained from the STRING database (Szklarczyk et al., 2019).

**Results**

We generated 4,641 genetic variants from 958,178 SNPs and Hi-C data and identified the transcriptomic network over a set of 1,181 transcripts from a class of genes that includes the highest number of SCZ-associated genes according to Fromer et al. (2016). Among 4,641 genomic variants, 81 showed significant effect on gene transcription level. The distributions of degrees (i.e. the sum of in- and out-degrees) for the genes influenced and for those not influenced by any genetic variants in the network are depicted in Figure 1B. Both histograms show similar distribution pattern, but the distribution without genetic variant is shifted towards degree $\geq 1$. Therefore, we conjectured that genes with high connectivity are less likely to be influenced by genomic variants (*cis*-regulation), thus their transcription is rather regulated directly or indirectly by other genes, mostly from different chromosomes (*trans*-regulation). Supplementary Table 1 lists the genes mapped influenced by genetic variants, together with in-/out-degrees.

We mapped the SCZ-associated genes on the identified transcriptomic network (blue colored nodes in Figure 2A), which revealed underlying relationships (pathways) among genes. To extract information from the transcriptomic causal networks, we measured the network properties and classified the genes as receptor, broadcaster, or mediators summarized in Figure 2C for SCZ-associated genes. Figure 2D clustered the SCZ associated genes based on type of studies that previously found the association (more details in Supplementary Table 2).

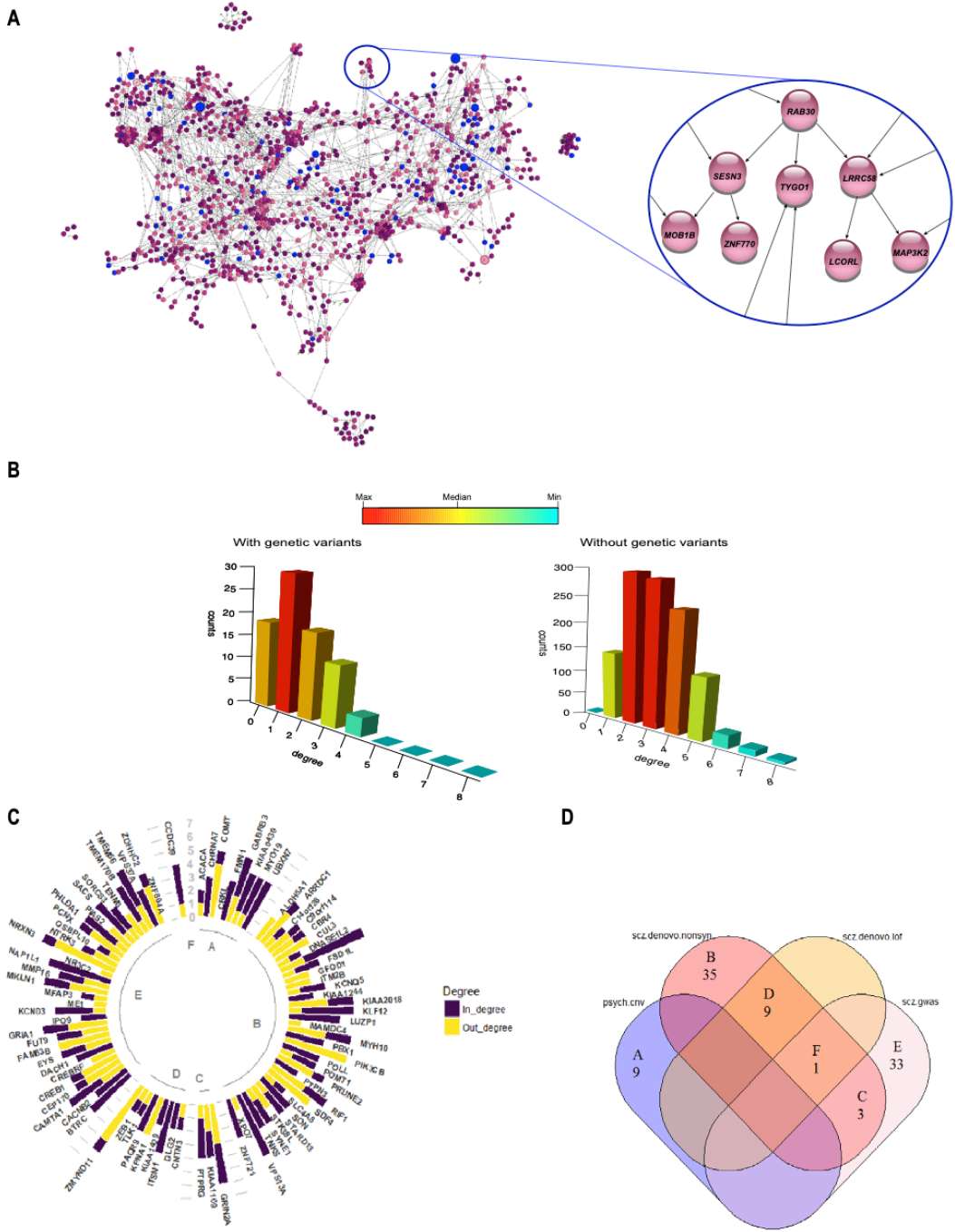

**Figure 2**. **A**: Identified transcriptomic causal network and a closeup. Each node represents a gene. Genes associated with SCZ are depicted in blue. **B**: Distribution of the connectivity of the genes with genetic variants (left) and the genes without genetic variants (right). **C**: Represents the in-degree and out-degrees of genes associated with SCZ in the network. Single color bars in yellow or purple are corresponding to broadcaster and receptor genes respectively while bars with both colors are representing mediator genes. **D**: Venn diagram cluster SCZ-associated genes based on the type of studies that previously found their association (psych.cnv: Psychiatric study of copy number variation; SCZ.denovo.nonsyn: SCZ study of non-synonymous de novo mutations; SCZ.gwas: SCZ study of genome wide association studies; SCZ.denova.lof: SCZ study of Loss-of-function de novo mutations).

## Differential gene regulatory patterns

### 1- Loss of mediators

We identified 9 mediator genes that lost their interactions with downstream genes in SCZ-cases as compared to controls. Among these, 3 genes (*GABRA2*, *LRRTM2*, *PPM1E*) have biased expression in human brain with no or minimal expression in other human tissues (Supplementary Table 3). Figure 3A shows the exclusive effect of these three gene on their downstream genes in the network. The subnetworks of these mediators are depicted in Figure 3B which shows the calculated causal effect size of the transcription level on their downstream genes.

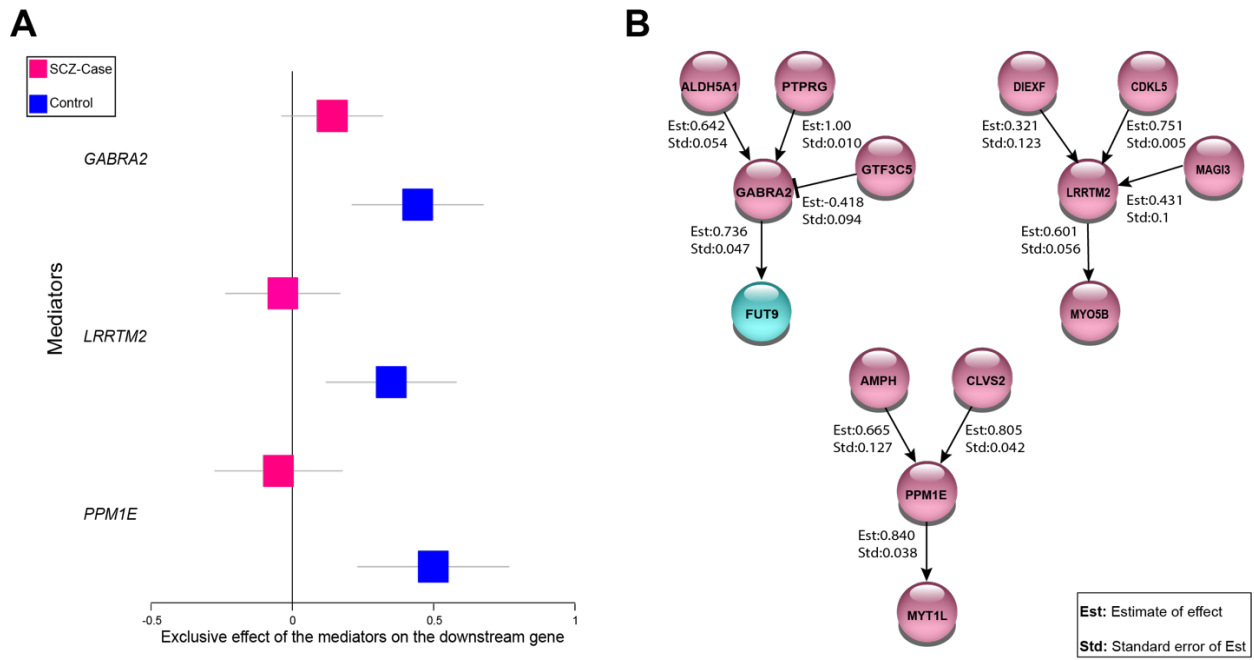

**Figure 3**. **A**: Effect of the mediators exclusively expressed in brain on the downstream genes. The forest plot represents 95% confidence intervals of the effect size in downstream gene transcription by the different mediator gene, in red mean effect size for SCZ cases and in blue, mean effect size for controls. **B**: Upstream genes and downstream genes for the mediators (*GABRA2, LRRTM2, PPM1E*) and their causal effects. The arrows represent transcription activation (positive effect size). The line with capped end represents transcriptional repression (negative effect size). In cyan is the gene reported as SCZ-associated gene in previous studies. For all interactions, the effect size (Est) and standard deviation (Std) are shown.

Using the STRING database (Szklarczyk et al., 2019), we found that the corresponding proteins GABRA2, LRRTM2, and PPM1E interact indirectly with the proteins encoded by the downstream genes in our network (Supplementary Table 4). Therefore, we believe that these proteins are extremely important for the normal brain function, since their corresponding genes (mediators) lost their effect on the downstream gene transcriptional regulation in SCZ-cases versus controls.

Among the other 6 genes with loss of mediators (Figure 4), *SORT1* and *GNAL* have the highest expression in brain; *ZNF692* and *RALGPS2* have highest expression in testis; ZNF672 and SNRNP48 are expressed in all tissues including brain. The corresponding proteins produced by all these genes interact directly/indirectly with the proteins encoded by the downstream genes in our network over control group (Supplementary Table 3). Therefore, based on our analysis, we believe these genes are implicated in SCZ, since these genes (mediators) lost their impact on the downstream gene transcriptional regulation in SCZ-cases versus controls.

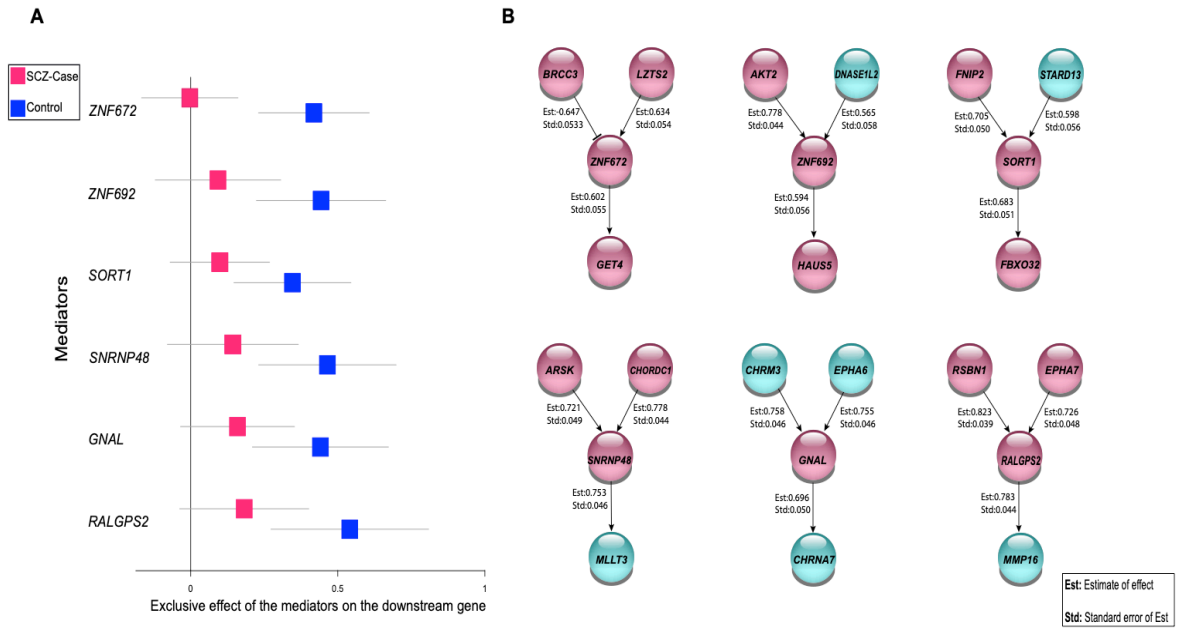

**Figure 4**. **A**: Effect of the mediator genes essential for normal brain functioning on the downstream genes. The forest plot represents 95% confidence intervals of the effect size in downstream gene transcription by the different mediator gene, in red mean effect size for SCZ cases and in blue, mean effect size for controls. **B**: Upstream genes and downstream genes for the mediators (*ZNF672, ZNF692, SORT1, SNRNP48, GNAL, RALGPS2*) and their causal effects. The arrows represent transcription activation (positive effect size). The line with capped end represents transcriptional repression (negative effect size). In cyan there are the genes reported as associated with SCZ in previous studies. For all interactions, the effect size (Est) and standard deviation (Std) are shown.

## 2- Modules

We identified several modules and predicted the amount of transcription of non-hub genes from the transcription level of the corresponding hub in both groups, SCZ cases and controls. In the following subsections, we described the modules summarized in Table 1 and depicted the modules as a sub-network in the corresponding figures for simplicity. For each module, the strength of the regulatory interactions in the network is represented by a heatmap of their corresponding p-values above a certain cutoff value, that is defined as the significance level of the likelihood for an edge to connect two genes (regulatory interaction) given the other genes in the network. Moreover, we discovered several differential regulatory patterns and two new genes putatively related to SCZ (Table 2) using prediction and conditional analysis.

Table 1. Summary of the main features of the identified modules named after the gene acting as the hub; the number of genes in the module, the number of SCZ-associated genes, the number of genes influenced by the hub in the longest path, and the Gene Ontology annotation.

| Module number | Hub gene | Number of genes | Number of SCZ associated genes | Number of genes in longest path | Gene Ontology |
|---|---|---|---|---|---|
| 1 | *TENM3* | 10 | 7 | 4 | Cellular Movement, reproductive system development & function, cardiovascular system development & function |
| 2 | *NRXN3* | 6 | 3 | 1 | Gene Transcription, RNA damage & repair, RNA post-transcriptional modification |
| 3 | *MYH10* | 8 | 5 | 3 | cellular development, cellular growth & proliferation, nervous system development & function |
| 3 | *PEX5L* | 8 | 2 | 4 | Cellular development, cellular Growth & proliferation, nervous system development & function |

Table 2. Novel SCZ associated genes and essential genes for brain functioning identified through regression and conditional analysis.

| Hypotheses | | | |
|---|---|---|---|
| Novel SCZ-associated genes | | Essential genes for brain function | |
| Genes | Modules | Essential gene | SCZ-associated genes controlled by essential genes |
| *SEZ6L, RNF150* | *TENM3*-Module | *TENM3* | *BTRC, RNF150, SEZ6L, GRIN3A, VAT1L, GRIA1, XKR4* |
| *UNC5D* | *NRXN3*-Module | *NRXN3* | *CNTNAP2, ARFGEF3* |
| *RTF1* | *MYH10*-Module | *MYH10* | *ANK2, LMO7, LRRC4C, DLG2, RTF1, PTPRK* |
| *FBXO32* | *PEX5L*-Module | *PEX5L* | *DLG2, STARD13* |

*TENM3*-**Module:** *TENM3* had previously associated to SCZ with impact on the transcription of 9 other genes downstream in our network (Supplementary, Table 5), 6 of which were known to be related to SCZ (Figure 5A, 5C). We explored the impact of *TENM3*, considered as the hub for this module, on non-hub genes within the module by predicting their transcription levels from *TENM3* using cross-validation approach. Figure 5A represents the module sub-network in controls. The prediction analysis resulted in a good performance in both SCZ and control groups (Figure 5B). Although *TENM3* predicts slightly better the transcription level for the genes in the control as compared to the SCZ group, the result supports the interaction of hub with non-hub genes in both cases and controls. Therefore, *TENM3* plays a key role at controlling the transcription level of *BTRC*, *RNF150*, *SEZ6L*, *GRIN3A*, *VAT1L*, *GRIA1*, and *XKR4* SCZ-associated genes.

The longest path in this module is *TENM3*→ *GRIN3A*→ *GRIA1*→ *SORCS3*→ *RNF150*. All the genes in this pathway are associated with SCZ except *RNF150*. Therefore, we hypothesized that *RNF150* is a SCZ-associated gene as well. This hypothesis is supported by the fact that another SCZ-associated gene from out of the module, FLRT2, directly affects *RNF150* transcription. We predicted *RNF150* transcription level based on the transcription of all three genes (*SORCS3*, *FLRT2*, *SEZ6L*) with direct impact on *RNF150* using a penalization-regression model and cross-validation (see methods for details). These three genes explained more than 70% of variation in *RNF150* in both groups (Figure 5D) which agrees with our hypothesis since one of the two upstream genes of *SEZ6L* is SCZ-associated gene as well. Therefore, *RNF150* can be a good candidate to study SCZ.

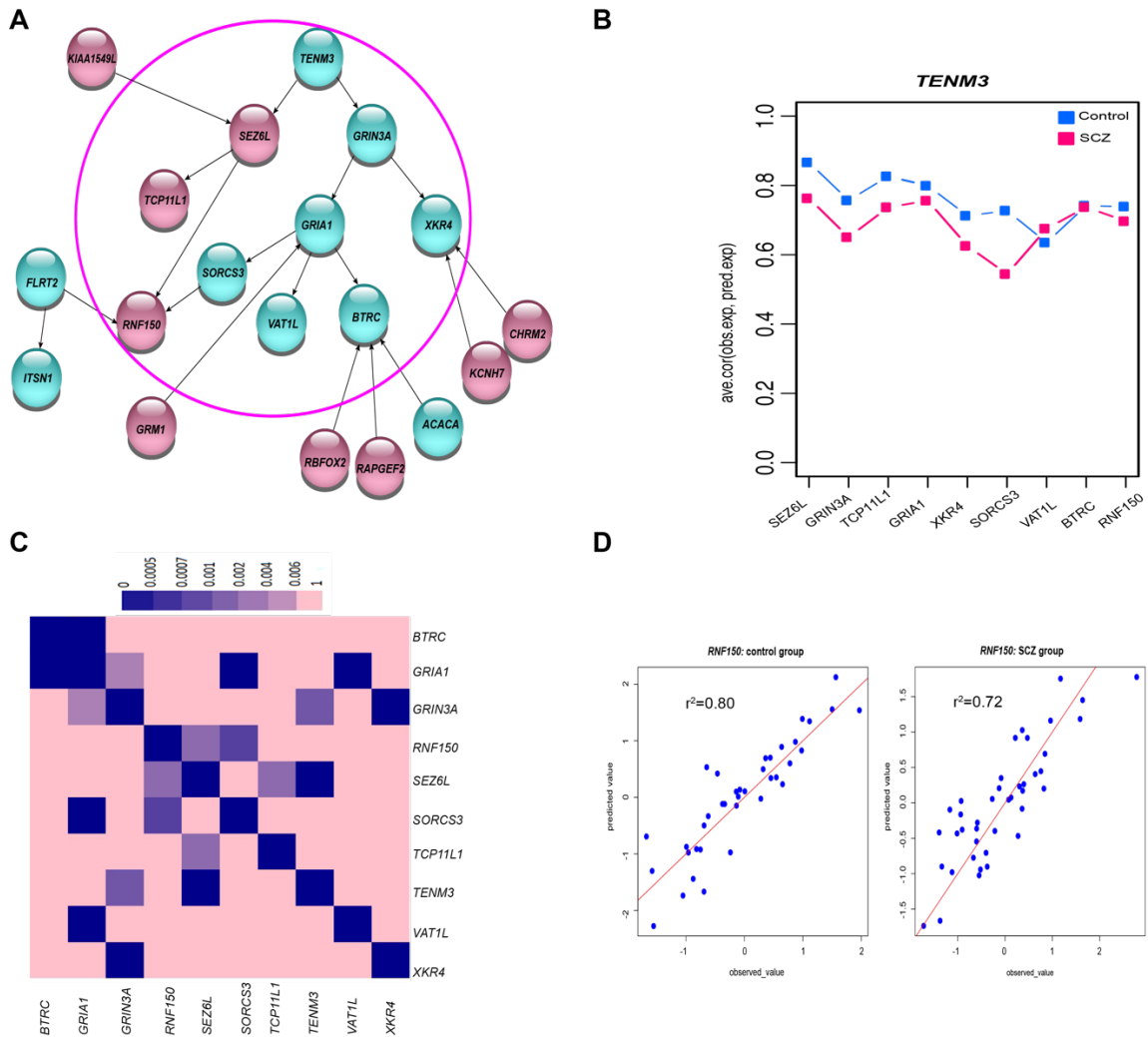

**Figure 5**. **A**: *TENM3*-Module with SCZ-associated genes in cyan. **B**: Prediction of non-hub gene transcription levels based on transcription of the hub. This plot shows the average Pearson correlation coefficient between observed and predicted values for controls (blue) and SCZ individuals (red). **C**: Heatmap of the strength of the relationship between genes in the modules based

on a p-value cut-off. **D**: Prediction of *RNF150* transcription level in both SCZ and control groups using *SEZ6L, FLRT2, SORCS3* transcription level as predictor.

We used the Human Protein Atlas (HPA) to see the tissue-specific expression of genes in this module, both at mRNA and protein levels (Supplementary Figure 1). All the genes are expressed to some extent in human brain, although for some of them (*SEZ6L*, GRIN3A, *XKR4* and *SORCS3*) there was no data available regarding the protein levels. The hub of the module, *TENM3,* showed high transcription and low protein level in human brain. It affected directly the transcription of two genes that are also highly transcribed in the brain (*SEZ5L* and *GRIN3A*) but no data is available for protein level. A search into the STRING database for physical interactions between the proteins coded by those genes with evidence of expression (GRIA1, BTRC, TCP11L1, RNF150) revealed the following interaction pairs: GRIA1– BTRC (meditated by DLG1), BTRC – ACACA (mediated by GSK3B) and BTRC – RAPGEF2 (mediated by CTNNB1). All these interactions are based on experimental evidences (see Supplementary Table 6).

*NRXN3*-**Module:** The NRXN3-Module (Figure 6A, C) includes 4 genes (*ARFGEF3*, *HECW2*, *DLGAP1*, *CNTNAP2*) whose transcription is directly influenced by *NRXN3* hub (Supplementary Table 7). Even though the effect of *NRXN3* itself does not propagate further within the network after one step, we believe it is an important gene since it encodes a member of a family of proteins that function in the nervous system as receptors and cell adhesion molecules [Ref]. *NRXN3* is preferentially transcribed in brain according to HPA and has also been found associated with SCZ in a Genome Wide Association Study (GWAS) (Hu et al., 2013). This module unveils that *MAS1* affects *NRXN3* and *FREM3* transcription, but the impact on *NRXN3* is much higher (Figure 6C). *NRXN3* exerts the effect of *MAS1* into the system although *FRM3* blocks the effect of *MAS1*. Interestingly, Gene Ontology analysis shows that all genes in the module except *FRM3* are related to "cancer, organismal injury and abnormalities, cellular development". Hence, we included *MAS1* in this module.

We predicted gene transcription levels in both SCZ, and control groups based on *NRXN3* transcription. Figure 6B shows the average of the Pearson correlation coefficients between predicted and observed transcription values for SCZ (red) and control (blue) groups. The good agreement between predictions of downstream transcription levels in both, SCZ cases and controls, is an evidence of the essential role of *NRXN3* in brain functioning.

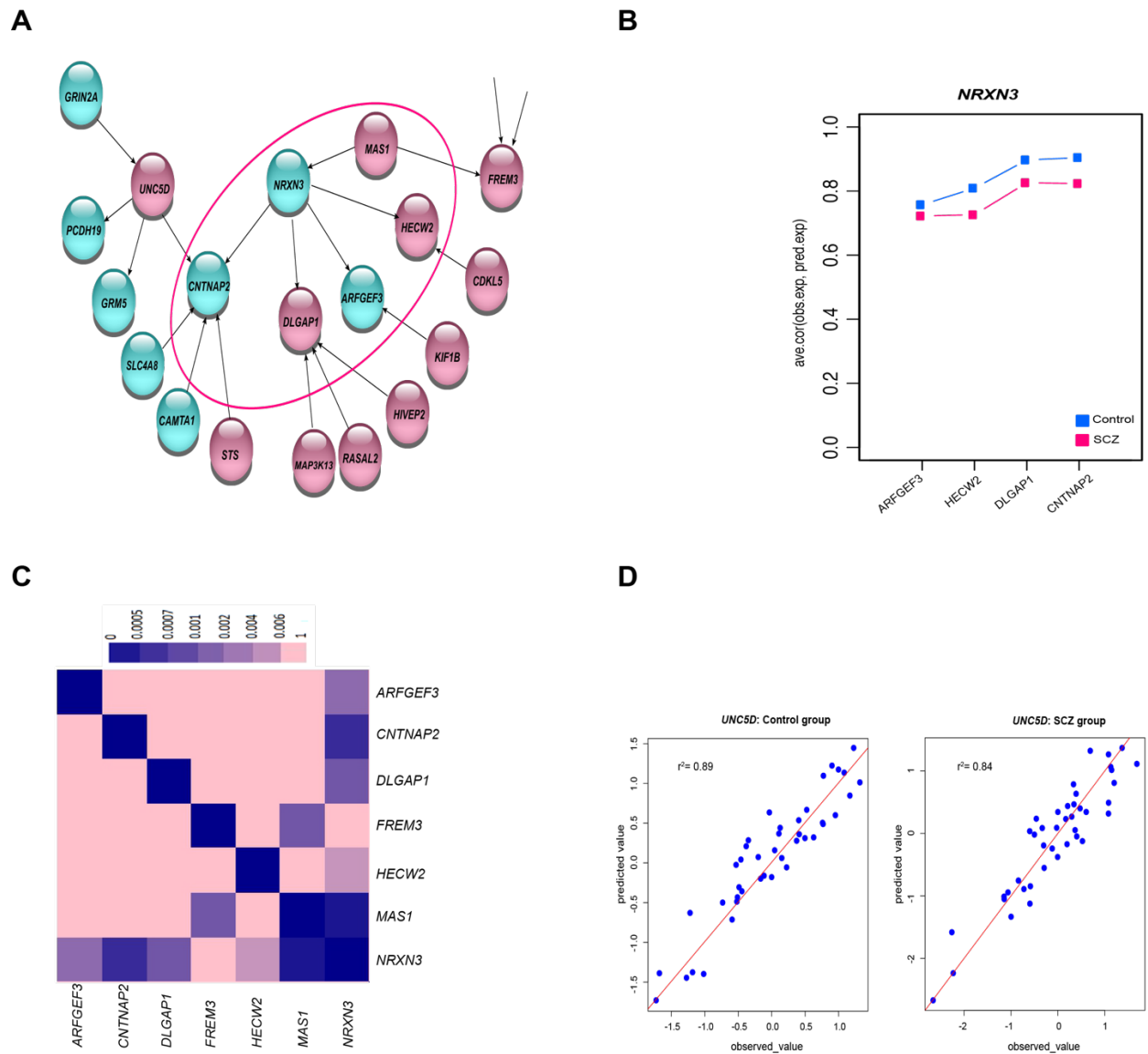

**Figure 6**. **A:** *NRXN3*-Module with previously SCZ-associated genes in cyan. **B**: Prediction of non-hub gene transcription levels based on transcription of the hub. The plot shows the average Pearson correlation coefficient between expected and predicted transcription. **C**: Heatmap of the strength of the relationship between genes in the modules based on the significance of their association in the network. **D**: Prediction *UNC5D* transcription level in both SCZ and control groups using *GRIN2A, PCDH19, GRM5, CNTNAP2* transcription levels.

The highest impact of *NRXN3* is on *CNTNAP2* gene (Figure 6A, C) which is influenced by 4 genes (*UNC5D*, *CNTNAP2*, *SLC4A8*, *CAMTA1* and *STS*) from outside the module. Among these 4 genes, *UNC5D* is the most relevant to our study, since its transcription directly influenced 3 SCZ-associated genes (*PCDH19, GRM5, CNTNAP2*) and is also influenced by a SCZ-associated gene (*GRIN2A*). Therefore, we strongly hypothesized that *UNC5D* is a gene related to SCZ given that *PCDH19, GRM5,* and *CNTNAP2* are associated with SCZ. We tested this hypothesis by predicting the transcription level of *UNC5D* as a linear outcome of the genes with direct connection (*GRIN2A,*

*PCDH19, GRM5, CNTNAP2*). As it can be seen in Figure 6D, the good predictions in both, cases and controls, supports our hypothesis.

According to the HPA, *NRXN3* is moderately expressed in the brain at both mRNA and protein levels (Supplementary Figure 2). Among 4 genes that are directly affected by the hub, *CNTNAP2* and *HECW2* show also high protein and mRNA levels in the brain. Although, there is no data available for *DLGAP1* and *ARFGEF3* at protein level in the, *DLGAP1* is one of the genes with highest mRNA level in the module. Searching for protein-protein interactions in STRING database, we found the following interactions: NRXN3 – DLGAP1; NRXN3 – CNTNAP2 (meditated by DLG); GRIN2A – PCHD19; GRIN2A – CNTNAP2; GRIN2A – GRM5. More details can be found in Supplementary Table 8.

**MYH10-Module**: *MYH10* is associated with SCZ that influences the transcription of 7 other genes in the network (Figure 7A and Supplementary Table 9) and its transcription is influenced by two upstream genes (*ADAM23* and *CHD6)*. In this module all the genes except those influenced by *MYH10* through *LMO7* pathway are associated to SCZ. *MYH10* and *LMO7* in turn, are in the longest and strongest path in this module (Figure 7A, C). The impact of the hub on the downstream genes are investigated using prediction model. The transcription of all target genes in the module can be predicted from the transcription activity of the *MHYH10* in SCZ-cases and controls, except for the furthest-related (*HMG20A*) gene to the hub (Figure 7B). Consequently, we hypothesized that *LMO7* and *RTF1* are also associated with SCZ. We assessed the effect of *LMO7* on *RTF1* to see if this influence is due to *MYH10* that is a SCZ-associated gene or on the contrary, it is due to *ROCK2* gene. Using conditional analysis, we calculated the exclusive effect in regard to *MYH10*, *ROCK2* and both (Figure 7D). Hence, *LOM7* and *RTF1* both are associated with SCZ. Furthermore, we conclude that *MYH10* has a key role in regulating the transcription of *ANK2, LMO7, LRRC4C, DLG2, RTF1, PTPRK* genes.

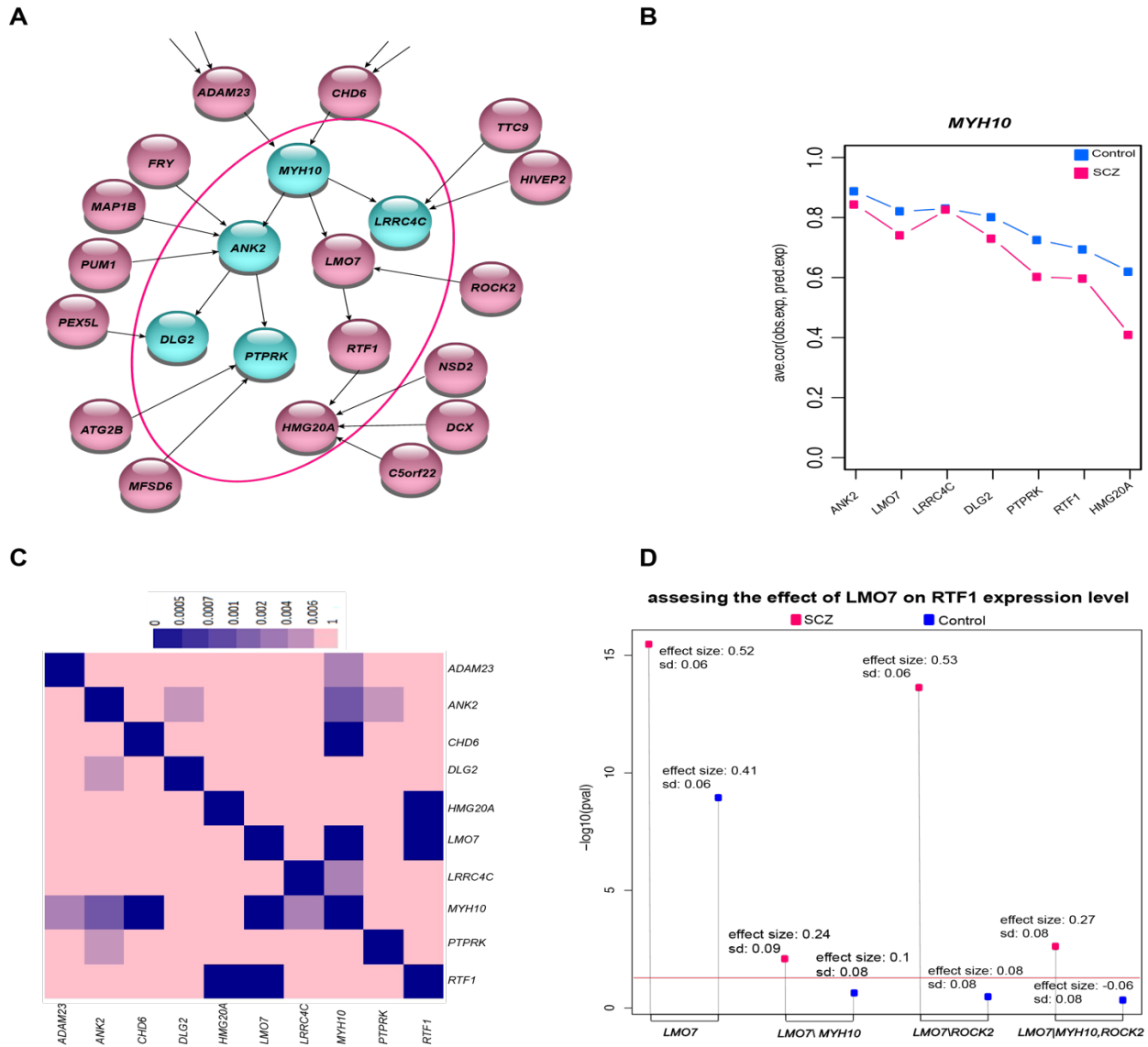

**Figure 7**. **A**: *MYH10*-Module with SCZ-associated genes in cyan. **B**: Prediction of non-hub gene transcription levels based on transcription of the hub. The plot shows the average Pearson correlation coefficient between expected and predicted transcription. **C**: Heatmap of the strength of the relationship between genes in the modules based on the significance of their association in the network. **D**: Summary-result of assessing the effect of *LMO7* on *RTF1* using conditional analysis in 4 different scenarios.

All the genes in this module are expressed significantly in brain, at both mRNA and protein levels (Supplementary, Table 10). STRING database showed the protein interactions of MYH10 - ANK2 (mediated by MYH9 and SPTAN1), ANK2 - DLG2 (mediated by NRCAM or NFASC), LRRC4C - DLG2 and MYH10 - RTF1 (mediated by MYH9 and CDC5L). MYH10 protein and its upstream donor in our network ADAM23, which shows medium protein level in brain according to HPA, interact through MYL3 mediator (Supplementary, Table 11). In addition, gene ontology analysis categorized all genes in *MYH10*-Module with similar Function, "cell-to-cell signaling and interaction, nervous system development and function, cell morphology".

**PEX5L-Module**: *PEX5L* has strong significant impact on two genes associated with SCZ (*STARD13* and *DLG2*) (Figure 8A, C and Table 12 in Supplementary) making *PEX5L* a good target for SCZ studies. The effect of *PEX5L* is very strong upon the genes of the first interaction shell and it decreases along the path through successive indirect interactions (Figure 8C). *PEX5L* lost its indirect effect on *FBXO32, ZBTB43,* and *TGOLN2* genes in SCZ-cases as compared to controls since it shows poor prediction performance for these genes for SCZ-cases (Figure 8B).

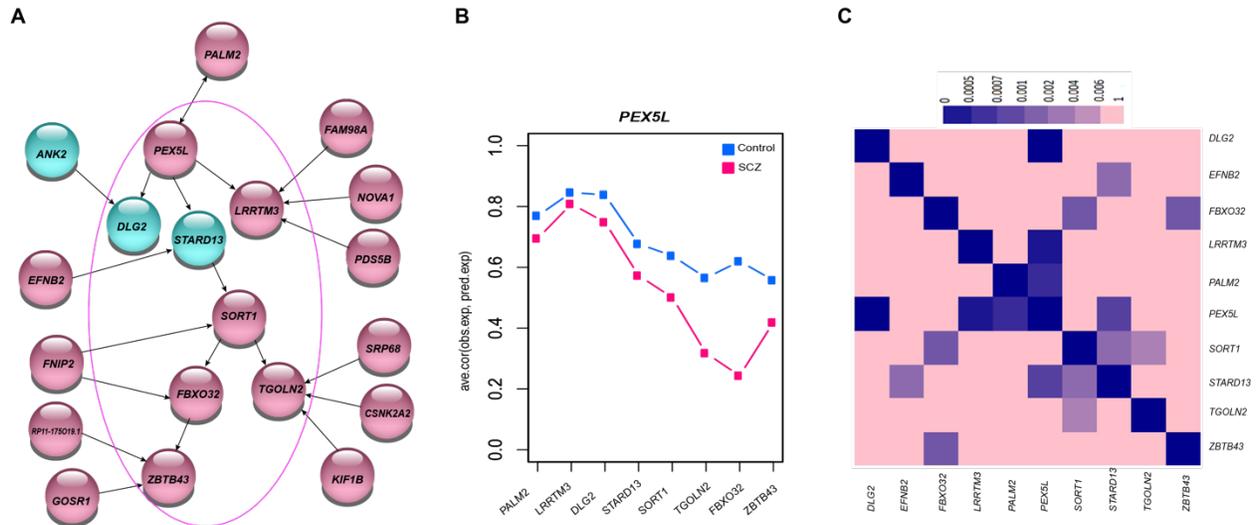

**Figure 8**. **A**: *PEX5L*-Module with a SCZ-associated in cyan. **B**: Prediction of non-hub gene transcription levels based on transcription of the hub. The plot shows the average Pearson correlation coefficient between expected and predicted transcription. **C**: Heatmap of the strength of the relationship between genes in the modules based on the significance of their association in the network.

*DLG2* is a receptor at the edge of the *PEX5L*- and *MYH10*-Modules. Although the strength of connection in the former module is higher than the latter, both hubs are good predictors for *DLG2* transcription level. We used protein-protein interaction data from STRING database and observed that DLG2 has stronger interaction with MYH10 protein through ANK2 (Supplementary Table 13). Therefore, we consider *DGL2* belongs to the *MYH10*-Module at the protein – protein interaction level, but to both modules at transcriptional regulation level.

Figure 8A also shows that the direction of *PALM2* interaction over *PEX5L* could not be identified through this analysis. While *PALM2* is out of the module, at the edge of the absolute network and has no genetic variants associated with it, we believe that *PALM2* transcription may be regulated by other genes that are not included in this study. All genes in this module share the same function related to "Cellular development, cellular Growth and proliferation, nervous system development and function" according to the Gene Ontology.

**Discussion:**

The effect of gene expression profile alterations on human health require to be investigated in relation with other genes in a particular tissue. In this study, we focused on understanding the gene regulatory pattern in human brain in the context of schizophrenia (SCZ). We introduced a novel approach for building transcriptomic causal network and for comparing the gene regulatory pattern between SCZ cases and controls. The transcription of a gene may be affected by genetic variants in *cis*-regulatory regions, but it can also be mediated by transcribed genes in *trans*. Using these observations, we integrated exome chip, Hi-C, RNA-seq data and built the transcriptional-causal network. Interestingly, genes with fewer connections are more likely to be affected by genetic variants than those highly connected in our network, but the latter are more often found regulated by or regulating other genes in *trans*. We also identified essential genes for brain function, whose transcription regulation does not change between cases and controls, using linear regression and cross-validation approaches. Additionally, we identified genes that are differentially regulated in SCZ cases.

Three genes, *GABRA2*, *LRRTM2* and *PPM1E* that are differentially regulated in individuals diagnosed with SCZ are also preferentially expressed in human brain. All these genes are related to synapse, which have been hypothesized to underlie altered neuronal function in complex neuropsychiatric disorders (Wang, Christian, Song, & Ming, 2018). In particular, *GABRA2* is believed to be a component of the hetero-pentameric receptor for GABA, the major inhibitory neurotransmitter in the brain. Consequently, it plays an important role in the formation of functional inhibitory GABAergic synapses in addition to mediating synaptic inhibition as a GABA-gated ion channel (based on homology). *LRRTM2* is involved in the development and maintenance of the excitatory synapse in vertebrates, regulates surface expression of AMPA receptors, and instructs the development of functional glutamate release sites. *PPM1E* encodes protein phosphatase that inactivates multifunctional CaM kinases, whose activation can affect many downstream pathways controlling a variety of cellular functions (Swulius & Waxham, 2008).

Our analysis shows differential regulation of *GNAL* between SCZ and control groups. In our network, *GNAL* that encodes a stimulatory G protein alpha subunit is a mediator between two SCZ drug targets (*CHRM3* and CHRNA7). The protein coded by the upstream gene *CHRM3* is targeted by antipsychotic drugs against schizophrenia and bipolar disorder but causes insulin dysregulation that may precede diabetes. *CHRNA7* as downstream gene of *GNAL* is also considered a promising drug target for treatment of cognitive dysfunction in schizophrenia and improves memory and executive functions in patients and healthy individuals. However, clinical trials with pro-cognitive drugs are challenged by large inter-individual response variations. This differential regulatory pattern suggests that *GNAL* can be an alternative SCZ drug target since it encodes a G protein alpha subunit that is widely expressed in the central nervous system.

The following four genes *PEX5L, TENM3, NRXN3, MYH10* are essential for brain function since they have high impacts on several genes in both SCZ and control groups. *PEX5L* has biased expression in brain. The corresponding PEX5L protein is believed to work as accessory subunit of hyperpolarization-activated cyclic nucleotide-gated channels, regulating their cell-surface expression and cAMP dependence (The UniProt Consortium, 2017). *PEX5L* exerts a significant effect directly on the transcription of three genes: *DGL2, STARD13* and *LRRTM3*. While the first two had been previously linked to SCZ, *LRRTM3* has also biased expression in brain. Homologous proteins to LRRTM3 are involved in the development and maintenance of the vertebrate nervous system (The UniProt Consortium, 2017). Furthermore, among all downstream genes of *PEX5L*, *FBXO32* shows the highest difference in regulation between cases and controls. This differential pattern is supported by the fact that FBXO32 promotes neuronal protein homeostasis through coordinating autophagy/lysosome-mediated protein turnover with the ubiquitin–proteasome system (Murdoch et al., 2016).

*TENM3* encodes a protein that translocates in the nucleus, regulating the transcriptional activity of genes related to neurite growth and cell adhesion [wiki from pfam]. This is clearly in agreement with our results since *TENM3* strongly affects multiple genes in our network. In addition, it has been shown that its homologous protein establishes the neuron connectivity in the mouse brain (Berns, DeNardo, Pederick, & Luo, 2018). NRXN3 protein is involved in synaptic plasticity (Kelai et al., 2008). Moreover, among the four genes directly influenced by *NRXN3*, three of them *(DLGAP1, DLGAP1, CNTNAP2)* are related to synaptic plasticity as well (Toro et al., 2010; Ullman, Smith-Hicks, Desai, & Stafstrom, 2018). Thus, it makes sense to consider all five genes associated with synaptic dysfunction, which is a causal factor for neuropsychiatric disorders including SCZ. *MYH10* gene is important for the normal development and function of dendritic spines ( Zhang, Webb, Asmussen, Niu, & Horwitz, 2005) (Rex et al., 2010)(Ryu et al., 2006)(Hodges, Newell-Litwa, Asmussen, Vicente-Manzanares, & Horwitz, 2011).

We conclude that genetic variation in the context of schizophrenia works as *cis*-regulatory elements for genes that are not so much involved in *trans*-regulatory interactions. This property may just be general of gene regulatory networks. Our method provided the regulatory context for essential genes in the brain. Besides, the proposed approach was also able to generate new sets of genes putatively associated with SCZ that are differentially regulated in cases. To the best of our knowledge, the latter will allow for the first time, the design of further experiments to target those relevant genes, either by editing them, or drug targeting, without disrupting essential pathways for normal brain function. Altogether this work presents a significant step towards the rational understanding of the genetic mechanisms of schizophrenia and pave the way for designing new and more selective treatments that minimize undesired secondary effects.